\documentclass[a4paper,twoside]{article}

\usepackage{epsfig}
\usepackage{subfigure}
\usepackage{calc}
\usepackage{amssymb}
\usepackage{amstext}
\usepackage{amsmath}
\usepackage{amsthm}
\usepackage{multicol}
\usepackage{pslatex}
\usepackage{apalike}
\usepackage{listings}

\usepackage[utf8]{inputenc}
\usepackage{url}
\usepackage{graphicx}
\usepackage{subfigure}
\usepackage{listings}
\usepackage{framed}

\usepackage{SCITEPRESS}     

\begin{document}
 \lstset{language=Pascal}
\title{Towards a Lightweight Multi-Cloud DSL for Elastic and Transferable Cloud-native Applications}
\author{\authorname{Peter-Christian Quint, Nane Kratzke}
\affiliation{L\"ubeck University of Applied Sciences, 23562 L\"ubeck, Germany\\Center of Excellence for Communication, Systems and Applications (CoSA)
}
\email{\{peter-christian.quint, nane.kratzke\}@fh-luebeck.de}}

\abstract{ 
 
 \begin{abstract}
  CCloud-native applications are intentionally designed for the cloud in order to leverage cloud platform features like horizontal scaling and elasticity -- benefits coming along with cloud platforms. In addition to classical (and very often static) multi-tier deployment scenarios, cloud-native applications are typically operated on much more complex but elastic infrastructures. Furthermore, there is a trend to use elastic container platforms like Kubernetes, Docker Swarm or Apache Mesos. However, especially multi-cloud use cases are astonishingly complex to handle. In consequence, cloud-native applications are prone to vendor lock-in. Very often TOSCA-based approaches are used to tackle this aspect. But, these application topology defining approaches are limited in supporting multi-cloud adaption of a cloud-native application at runtime. In this paper, we analyzed several approaches to define cloud-native applications being multi-cloud transferable at runtime. We have not found an approach that fully satisfies all of our requirements. Therefore we introduce a solution proposal that separates elastic platform definition from cloud application definition. We present first considerations for a domain specific language for application definition and demonstrate evaluation results on the platform level showing that a cloud-native application can be transfered between different cloud service providers like Azure and Google within minutes and without downtime. The evaluation covers public and private cloud service infrastructures provided by Amazon Web Services, Microsoft Azure, Google Compute Engine and OpenStack.
\end{abstract}}
\onecolumn \maketitle \normalsize \vfill

 \section{\uppercase{Introduction}}
\label{sec:introduction}

\noindent Elastic container platforms (ECP) like Docker Swarm, Kubernetes (k8s) and Apache Mesos received more and more attention by practitioners in recent years \cite{DeAlfonso2017} -- and this trend still seems to continue \cite{KQ2017}. Elastic container platforms fit very well with existing cloud-native application (CNA) architecture approaches \cite{KQ2017}. Corresponding system designs often follow a microservice-based architecture \cite{Sill2016,KP2016}. Nevertheless, the reader should be aware that the effective and elastic operation of such kind of elastic container platforms is still a question in research -- although there are interesting approaches making use of bare metal \cite{DeAlfonso2017} as well as public and private cloud infrastructures \cite{KQ2017}. What is more, there are open issues how to design, define and operate cloud applications on top of such container platforms pragmatically. This is especially true for multi-cloud contexts. Such open issues in scheduling microservices to the cloud come along with questions regarding interoperability, application topology and composition aspects \cite{Saatkamp2017} as well as elastic runtime adaption aspects of cloud-native applications \cite{Fazio2016}. The combination of these three aspects (multi-cloud interoperability, application topology definition/composition and elastic runtime adaption) is -- to the best of the authors' knowledge -- not solved satisfactorily so far. These three problems are often seen in isolation. In consequence, \textbf{topology based multi-cloud approaches often do not consider elastic runtime adaption} of deployments \cite{Saatkamp2017} and \textbf{multi-cloud capable elastic solutions being adaptive at runtime do not make use of topology based approaches} as well  \cite{KQ2017}. And finally, (topology-based) cloud-native applications making use of elastic runtime adaption are often inherently bound to specific cloud infrastructure services (like cloud provider specific monitoring, scaling and messaging services) making it hard to transfer these cloud applications easily to another cloud provider or even operate them across providers at the same time \cite{KP2016}. Furthermore, Heinrich et al. mention several research challenges and directions like microservice focused performance monitoring under runtime adaption approaches \cite{Heinrich2017}. All in all, it seems like cloud engineers (and researchers as well) just trust in picking only two out of three options. \textit{Is this really the best approach?}
    
Therefore, this contribution strives for a more integrated point of view to overcome the observable isolation of these mentioned engineering and research trends \cite{KQ2017} and tries to analyze how and whether the mentioned approaches can be combined. We intentionally strive for a pragmatic and practitioner acceptance instead of richness of expression like this is done by approaches like CAMEL \cite{rossini2015cloud}. Other than CAMEL, we focus microservice architectures and elastic container platforms only to reduce language complexity. So, we do not follow an holistic approach considering every imaginable architectural style of cloud applications.

We present a prototype for a domain-specific language (DSL) that enables to describe cloud-native applications being transferable at runtime without downtimes according to the following \textbf{outline}. The key idea is to describe the platform independently from the application. Our DSL has been developed according to a three step DSL design methodology: analysis, implementation and use as proposed by \cite{van2000domain,Mernik2005,Strembeck2009}. In Section \ref{sec:ecp_based_cna} we analyze common characteristics of elastic container platforms and derive concepts that have to be covered by cloud-native application definition DSLs.  In Section \ref{sec:dsl} we refine these concepts into more concrete requirements and analyze related work and existing DSLs like TOSCA. We found no existing language that fulfills all of our identified requirements completely. Accordingly, we propose and present a prototypic implementation for such a DSL and evaluate it in Section \ref{sec:evaluation}. Our evaluation shows, that cloud-native applications can be transferred between different cloud service providers like Azure and Google within minutes and without downtime. We have executed our experiments on public and private cloud service infrastructures provided by Amazon Web Services, Microsoft Azure, Google Compute Engine and OpenStack. This section closes with a critical discussion. Finally, we conclude our considerations and provide an outlook in Section \ref{sec:conclusion}.
 
 \begin{figure*}[t]
 	\begin{center}
 		\includegraphics[width=\textwidth]{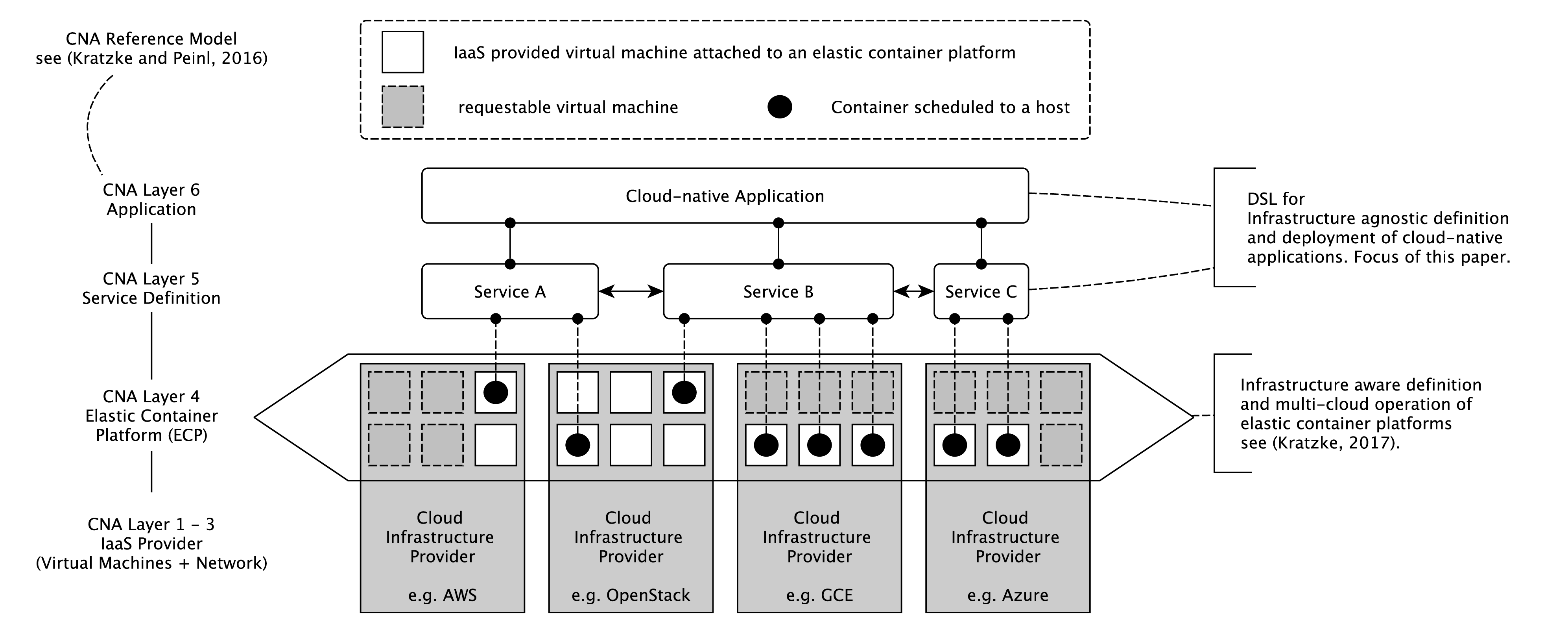}
 		\caption{An ECP based CNA deployment for elastic and multi-cloud capable operation.} 
 		\label{fig:dsl_layer}
 	\end{center}
 \end{figure*}
 
\section{\uppercase{Containerization Trends}} 
\label{sec:ecp_based_cna}

\noindent According to \cite{KQ2017}, a CNA runs on top of an elastic runtime environment. This can be straightaway an Infrastructure-as-a-Service (IaaS) or an elastic platform \cite{FLR+2014}. Container-based elastic platforms are getting more and more widespread. Such kind of elastic ECP are shown in Table \ref{tab:platforms}. For the aim to avoid vendor lock-in \cite{KP2016} we propose to make use of basic and standardized IaaS service concepts only. Such concepts are virtual machines, virtualized (block-)storage devices, virtualized networks and security groups. Elastic container platforms can be deployed on top of these basic IaaS service concepts \cite{Kratzke2017}. And on top of elastic container platforms arbitrary cloud-native applications can be deployed \cite{KP2016}.

Figure \ref{fig:dsl_layer} illustrates such kind of ECP based CNA deployments. \cite{Kratzke2017} showed that arbitrary ECPs can be operated using a descriptive cluster definition model based on an intended and a current state. Such kind of defined clusters can be operated or even transfered across different cloud service provider infrastructures at runtime. Obviously, this is a great foundation to avoid vendor lock-in situations. While a descriptive cluster definition model can be used for describing the elastic platform \cite{Kratzke2017}, there is also the need to describe the application topology without dependency to a specific ECP. This paper proposes to do this using a domain-specific language which focuses on the layer 5 and 6 of the cloud-native application reference model proposed by \cite{KP2016}. This DSL is the focal point of this paper. The central idea is to split the migration problem into two independent engineering problems which are too often solved together.

\begin{enumerate}
	\item The \textbf{infrastructure aware} deployment and operation of ECPs: These platforms can be deployed and operated in a way that they can be transferred across IaaS infrastructures of different private and public cloud services as \cite{Kratzke2017} showed.
	\item The \textbf{infrastructure agnostic} deployment of applications on top of these kind of transferable container platforms which is the focus of this paper.
\end{enumerate}

\begin{table}[t]
	\caption{\textbf{Some popular open source elastic platforms.} \textit{These kind of platforms can be used as a kind of cloud infrastructure unifying middleware.}}
	\label{tab:platforms} 
	\centering
	\footnotesize
	\begin{tabular}{lll}
		\hline
		\textbf{Platform} & \textbf{Contributors} & \textbf{URL}\\
		\hline
		Kubernetes & CNCF & \scriptsize \url{http://kubernetes.io} \\
		Swarm         & Docker & \scriptsize \url{https://docker.io} \\
		Mesos         & Apache & \scriptsize \url{http://mesos.apache.org} \\
		\hline
	\end{tabular}
\end{table}

In order to enable an ECP-based CNA deployment by a domain-specific language that is not bound to a specific ECP, the particular characteristics and commonalities of the target systems have to be identified. Therefore, the architectures and concepts of elastic container platforms have to be analyzed and compared. As representatives, we have chosen the three most often used elastic container platforms Kubernetes, Docker Swarm Mode and Apache Mesos with Marathon listed in Table \ref{tab:platforms}. As Table \ref{tab:ecpComp} shows all of these ECPs provide comparable concepts (from a bird's eye view).

\begin{table*}[t]
\caption{Concepts of analyzed ECP Architectures}
\label{tab:ecpComp}
\centering
\footnotesize
\begin{tabular}{llll}
\hline
\textbf{Concept}                     & \textbf{Mesos} & \textbf{Docker Swarm }                & \textbf{Kubernetes} \\
\hline
Application Definition & Application Group       & Compose                      & Service + Namespace \\
                                   &                                     &                                       & Controller (Deployment, DaemonSet, Job, ...) \\
                                   &                                     &                                      & All K8S concepts are described in YAML \\
\hline
Service discovery       & Mesos DNS                 & Service names & KubeDNS (or replacements) \\
                                   &                                      & Service links                & \\
\hline
Deployment Unit        & Binaries                        &                    Container (Docker)       & Pod (Docker, rkt) \\
                                   & Pods (Marathon)        &                                       & \\
\hline
Scheduling                 & Marathon Framework & Swarm scheduler         & kube-scheduler \\
                                   & Constraints                 & Constraints                   & Affinities + (Anti-)affinities\\
\hline
Load Balancing          & Marathon-lb-autoscale & Ingress load balancing & Ingress controller \\
                                   &                                     &                                      & kube-proxy \\
\hline
Autoscaling                & Marathon-autoscale    & -                                   & Horizontal pod autoscaling \\
\hline
Component Labeling           & key/value                      & key/value                      & key/value\\
\hline
\end{tabular}
\end{table*}

\textbf{Application Definition}. All platforms define applications as a set of deployment units. The dependencies of these deployment units are expressed in a descriptive way. Apache Mesos uses Application Groups to partition multiple applications into sets. The dependencies are modeled as n-ary trees of groups with applications as leaves. Kubernetes manages an application basically as a set of services composed of pods. A pod can contain one or more containers. All containers grouped in a pod run on the same machine in the cluster. ReplicationSet Controllers take care that the number of running pods is equal to the amount of pods defined in replication controller configurations \cite{Verma2015}. The numbers of running instances of a pod is defined in a so called deployment (YAML-file).
Docker Swarm supports application description using a single YAML-file that defines a multi-container deployment consisting of the container and there connections. YAML based definition formats seem to be common for all ECPs and \textit{a DSL should provide something like a model-to-model transformation (M2M) to these ECP specific application definition formats \textbf{[AD]}}.

\textbf{Service discovery} is the task to get service endpoints by name and not by a (permanently) changing address. All analyzed ECPs supported service discovery by DNS based solutions (Mesos, Kubernetes) or using the service names defined in the application definition format (Docker Swarm). \textit{Thus, a DSL must consider to name services in order to make them discoverable via DNS or ECP-specific naming services \textbf{[SD]}.}

\textbf{Deployment units}. The basic units of execution are named different by the ECPs. However, they mostly based on containers. Docker Swarm is intentionally designed for deploying Docker containers. A Kubernetes deployment unit is called a pod. And a pod whose the container can be operated by arbitrary container runtime environments. But Rocket (rkt) and Docker are the main container runtime environments at the time of writing this paper. Only Apache Mesos supports by its design arbitrary binaries as deployment units. However, the Marathon framework supports container workloads based on Docker containers and emerges as a standard for the Mesos platform to operate containerized workloads. \textit{A DSL should consider that the deployment unit concept (whether named application group, pod or container) is the basic unit of execution for all ECPs \textbf{[DU]}.}

\textbf{Scheduling}. All ECPs provide some kind of a scheduling service that mostly runs on the master nodes of these platforms. The scheduler assigns deployment units to nodes of the ECP considering the current workload and resource efficiency. The scheduling process of all ECPs can be constrained using scheduling constraints or so called (anti-)affinities \cite{Verma2015,naik2016building,hindman2011mesos}. \textit{These kind of scheduling constraints must be considered and expressable by a DSL \textbf{[SCHED]}.}

\textbf{Load Balancing}. Like scheduling, load balancing is supported by almost all analyzed platforms using special add-ons like Marathon-lb-autoscale (Mesos), kube-proxy (Kubernetes), or Ingress service (Docker Swarm). These load balancers provide basic round-robin load balancing strategies and they are used to distribute and forward IP traffic to the deployment units under execution. \textit{However, more sophisticated  load balancing strategies should be considered as future extensions for a DSL \textbf{[LB]}.}

\textbf{Autoscaling}. Except for Docker Swarm, all analyzed ECPs provide (basic) autoscaling features which rely mostly on measuring CPU or memory metrics. In case of Docker Swarm this could be extended using an add-on monitoring solution triggering Docker Compose file updates. The Mesos platform provides Marathon-autoscale for this purpose and Kubernetes relies on a horizontal pod autoscaler. Furthermore, Kubernetes supports even making use of custom metrics. \textit{So, a DSL should provide support for autoscaling supporting custom and even application specific metrics \textbf{[AS]}.}

\textbf{Component labeling}. All ECPs provide a $key/value$ based labeling concept that is used to name components (services, deployment units) of applications. This labeling is used more (Kubernetes) or less (Docker Swarm) intensively by concepts like service discovery, schedulers, load balancers and autoscalers. These concepts could be also of use for the operator of the cloud-native application to constrain scheduling decisions in multi-cloud scenarios. This component labeling can be used to code datacenter regions, prices, policies and even enable to deploy services only on specific nodes \cite{Kratzke2017}. \textit{In consequence, a DSL should be able to label application components in key/value style \textbf{[CL]}.}


\section{\uppercase{A DSL for CNA}}
\label{sec:dsl}

\begin{figure}[b]
	\begin{center}
		\includegraphics[width=\columnwidth]{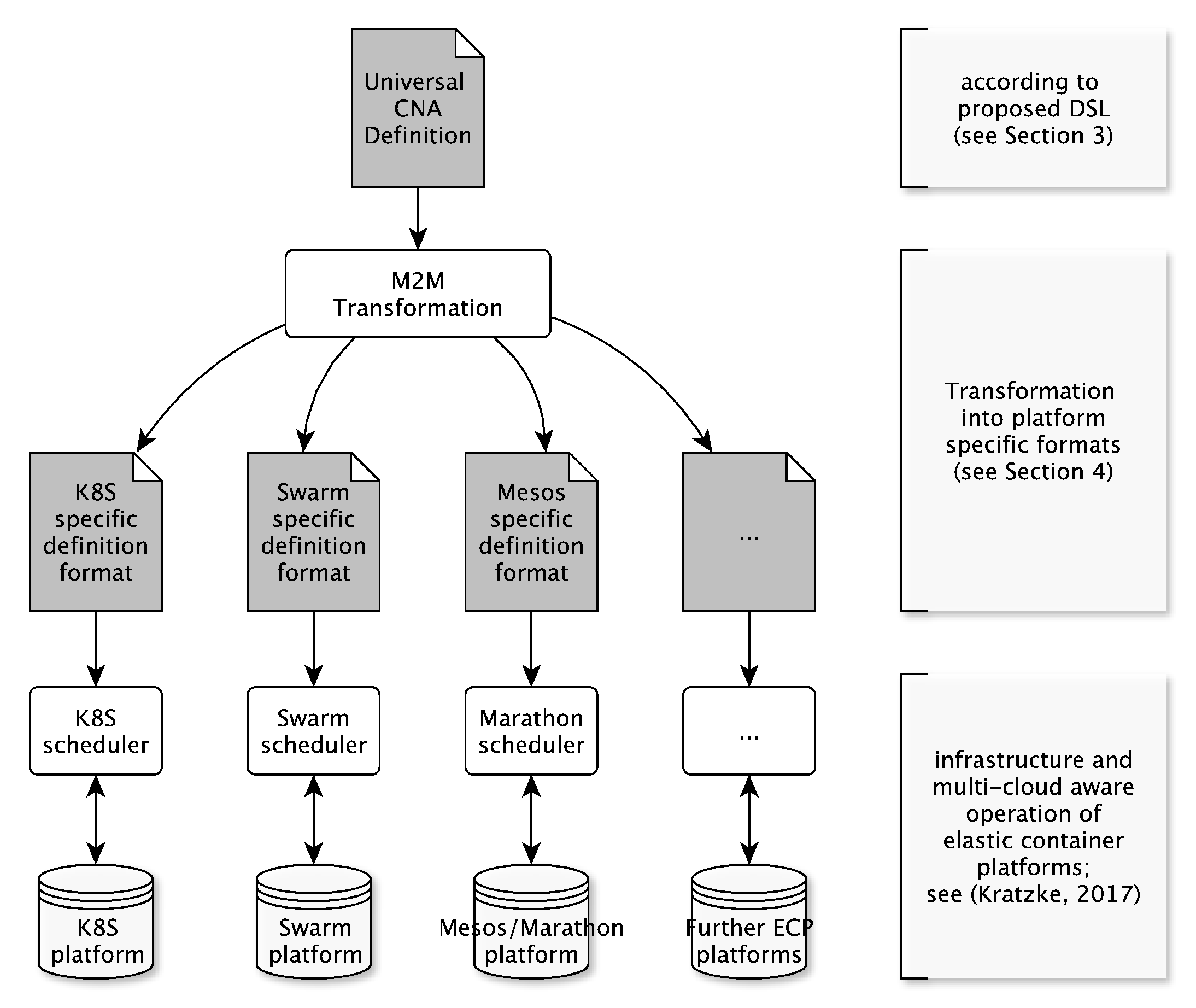}
		\caption{Separation of concerns in deploying a CNA} 
		\label{fig:dsl_flow}
	\end{center}
\end{figure}

\noindent For deploying arbitrary CNAs on specific elastic container platforms, we have developed a model-to-model (M2M) generator. As shown in Figure \ref{fig:dsl_flow}, the generated, ECP specific CNA description can be used by a ECP scheduler to deploy a new application or update a still running one. The operation of the ECPs can be done in a multi-cloud way \cite{Kratzke2017}. The elastic container platforms can be transferred across different cloud service platforms or they can be operated in a multi- or hybrid-cloud way. More details can be found in \cite{Kratzke2017}.

\begin{table*}[t]
\centering
\caption{Requirement Matching}
\label{tab:dslReq}
\resizebox{\textwidth}{!}{%
\begin{tabular}{ccccccccc}
\hline
\multicolumn{1}{|c|}{\textit{\textbf{Requirement \#}}} & \multicolumn{1}{c|}{\begin{tabular}[c]{@{}c@{}}\textbf{R1}\\Container\\$$\end{tabular}} & \multicolumn{1}{c|}{\begin{tabular}[c]{@{}c@{}}\textbf{R2}\\Application\\Scaling\end{tabular}} & \multicolumn{1}{c|}{\begin{tabular}[c]{@{}c@{}}\textbf{R3}\\Compendiously\\$ $\end{tabular}} & \multicolumn{1}{c|}{\begin{tabular}[c]{@{}c@{}}\textbf{R4}\\Multi-Cloud\\Support\end{tabular}} & \multicolumn{1}{c|}{\begin{tabular}[c]{@{}c@{}}\textbf{R5}\\Independence\\$ $\end{tabular}} & \multicolumn{1}{c|}{\begin{tabular}[c]{@{}c@{}}\textbf{R6}\\Elastic\\Runtime Env.\end{tabular}} & \multicolumn{1}{c|}{\textbf{Implementations}} & \multicolumn{1}{c|}{\textbf{Describtion}} \\ \hline
\multicolumn{1}{|c|}{\textbf{CAMEL}} & \multicolumn{1}{c|}{\textbf{}} & \multicolumn{1}{c|}{+} & \multicolumn{1}{c|}{-} & \multicolumn{1}{c|}{+} & \multicolumn{1}{c|}{+}  & \multicolumn{1}{c|}{+} &  \multicolumn{1}{c|}{R}&\multicolumn{1}{l|}{\begin{tabular}[l]{@{}l@{}}CAMEL is designed for modeling and execution of multi-cloud applications \cite{rossini2015cloud}.\\It integrates and extends existing DSLs like CloudML and supports models@run-time\\\cite{blair2009models},\cite{Chauvel2013}, an environment to provide a model-based representation\\of the underlying running system, which facilitates reasoning and adaptation of multi-cloud applications.\end{tabular}} \\ \hline
\multicolumn{1}{|c|}{\textbf{CAML}} & \multicolumn{1}{c|}{\textbf{}} & \multicolumn{1}{c|}{} & \multicolumn{1}{c|}{-} & \multicolumn{1}{c|}{-} & \multicolumn{1}{c|}{+}  & \multicolumn{1}{c|}{} &  \multicolumn{1}{c|}{R}&\multicolumn{1}{l|}{\begin{tabular}[l]{@{}l@{}}CAML enables the use of provider-dependent services (described in the CAML Profiles) and the\\deployment (described in the CAML Library). The cloud applications deployment\\configuration can be reused by using CAML templates \cite{bergmayr2014} \end{tabular}} \\ \hline
\multicolumn{1}{|c|}{\textbf{CloudML}} & \multicolumn{1}{c|}{\textbf{+}} & \multicolumn{1}{c|}{+} & \multicolumn{1}{c|}{-} & \multicolumn{1}{c|}{+} & \multicolumn{1}{c|}{+}  & \multicolumn{1}{c|}{+} & \multicolumn{1}{c|}{R} & \multicolumn{1}{l|}{\begin{tabular}[l]{@{}l@{}}A DSL for multi-cloud application deployments. The CloudMF \cite{lushpenko2015using}\\framework consists of CloudML \cite{Brandtzaeg2012} and models@run-time (see row above)\end{tabular}} \\ \hline
\multicolumn{1}{|c|}{\textbf{Docker Compose}} & \multicolumn{1}{c|}{+} & \multicolumn{1}{c|}{+} & \multicolumn{1}{c|}{+} & \multicolumn{1}{c|}{+} & \multicolumn{1}{c|}{-}  & \multicolumn{1}{c|}{+}  & \multicolumn{1}{c|}{P} & \multicolumn{1}{l|}{\begin{tabular}[l]{@{}l@{}}Is an orchestration DSL and tool for defining, linking, and running multiple containers\\on any docker host, also on the container cluster Docker Swarm\end{tabular}} \\ \hline
\multicolumn{1}{|c|}{\textbf{Kubernetes}} & \multicolumn{1}{c|}{+} & \multicolumn{1}{c|}{+} & \multicolumn{1}{c|}{+} & \multicolumn{1}{c|}{+} & \multicolumn{1}{c|}{-}   & \multicolumn{1}{c|}{+} &  \multicolumn{1}{c|}{P}  &  \multicolumn{1}{l|}{\begin{tabular}[l]{@{}l@{}}Kubernetes (former Google Borg) is a cluster platform for deploying container \\applications. All configurations like the scheduling units (pods) and the scaling properties\\(replication controller) can be described in YAML files \cite{Verma2015}\end{tabular}}\\ \hline\multicolumn{1}{|c|}{\textbf{TOSCA}} & \multicolumn{1}{c|}{+} & \multicolumn{1}{c|}{+} & \multicolumn{1}{c|}{-} & \multicolumn{1}{c|}{+} & \multicolumn{1}{c|}{+}  & \multicolumn{1}{c|}{-}  & \multicolumn{1}{c|}{R\&P}  & \multicolumn{1}{l|}{\begin{tabular}[l]{@{}l@{}}A specification for describing the topology and orchestration of cloud webservices, their \\relations and components of composition and how to manage them \cite{Binz2014}\end{tabular}} \\ \hline
\multicolumn{1}{|c|}{\textbf{MODACloudML}} & \multicolumn{1}{c|}{+} & \multicolumn{1}{c|}{+} & \multicolumn{1}{c|}{-} & \multicolumn{1}{c|}{+} & \multicolumn{1}{c|}{+}  & \multicolumn{1}{c|}{+} & \multicolumn{1}{c|}{R}& \multicolumn{1}{l|}{\begin{tabular}[l]{@{}l@{}}Designed to specify the provision and deployment of applications in multi-cloud\\environments \cite{artavc2016model}. MODACloudsML is the DSL part of MODACloud and\\also uses CloudMF (see row CloudML)\end{tabular}}  \\ \hline
\multicolumn{1}{|c|}{\textbf{MULTICLAPP}} & \multicolumn{1}{c|}{} & \multicolumn{1}{c|}{-} & \multicolumn{1}{c|}{} & \multicolumn{1}{c|}{+} & \multicolumn{1}{c|}{+} & \multicolumn{1}{c|}{-} & \multicolumn{1}{c|}{-} &\multicolumn{1}{l|}{\begin{tabular}[l]{@{}l@{}}A framework for modeling cloud applications on multi-cloud environments,\\independent from the IaaS-provider. Applications can be modeled with an UML profile \end{tabular}} \\ \hline
 \multicolumn{8}{l}{Legend for column \textit{Implementation}: P: Productive useable implementations available;  R: Research implementations available}
\end{tabular}%
}
\end{table*}

To define a universal CNA definition DSL, we followed established methodologies for DSL development as proposed by \cite{van2000domain,Mernik2005,Strembeck2009}. According to Section \ref{sec:ecp_based_cna} the following requirements arise for a DSL with the intended purpose to define elastic, transferable, multi-cloud-aware cloud-native applications being operated on elastic container platforms.

\begin{footnotesize}
\begin{itemize}
	\item \textbf{R1: Containerized deployments}. Containers are self-contained deployment units of a service and the core building block of every modern cloud-native application. \textbf{The DSL must be designed to describe and label a containerized deployment of discoverable services}. This requirement comprises \textbf{[SD], [DU], and [CL]}. 
	\item \textbf{R2: Application Scaling}.	Elasticity and scalability are one of the major advantages using cloud computing \cite{Vaquero2011}. Scalability enables to follow workloads by request stimuli in order to improve resource efficiency \cite{Mao2011}. \textbf{The DSL must be designed to describe elastic services.} This requirement comprises \textbf{[SCHED], [LB], and [AS]}.
	\item \textbf{R3: Compendiously}. To simplify operations the DSL should be pragmatic. Our approach is based on a separation between the description of the application and the elastic container platform. \textbf{The DSL must be designed to be lightweight and infrastructure-agnostic.} This requirement comprises \textbf{[AD], [SD], and [CL]}.
	\item \textbf{R4: Multi-Cloud-Support}. Using multi-cloud-capable ECPs for deploying CNAs is a major requirement for our migration approach. Multi-cloud support also enables the use of Hybrid-cloud infrastructures. \textbf{The DSL must be designed to support multi-cloud operations}. This requirement comprises \textbf{[SCHED], [CL]} and the necessity to be applied on ECPs operated in a way described by \cite{Kratzke2017}.
	\item \textbf{R5: Independence}.	To avoid dependencies, the CNA should be deployable independently to a specific ECP and also to specific IaaS providers. \textbf{The DSL must be designed to be independent from a specific ECP or cloud infrastructure.} This requirement comprises \textbf{[AD]} and the necessity to be applied on ECPs operated in a way described by \cite{Kratzke2017}.
	\item \textbf{R6: Elastic Runtime Environment}. Our approach provides a CNA deployment on an ECP which is transferable across multiple IaaS cloud infrastructures. \textbf{The DSL must be designed to define applications being able to be operated on an elastic runtime environment}. This requirement comprises \textbf{[SD], [SCHED], [LB], [AS], [CL]} and should consider the operation of ECPs in way that is described in \cite{Kratzke2017} .
\end{itemize}
\end{footnotesize}

According to these requirements, we examined existing domain-specific languages for similar kind of purposes. By investigating literature and conducting practical experiments in expressing a reference application, we analyzed whether the DSL fulfills our requirements. The results  are shown in Table \ref{tab:dslReq}. No of the examined DSLs covered all of the requirements. TOSCA, Docker Compose and Kubernetes fulfill the most of our requirements. But Docker Compose and the Kubernetes DSL are designed for a specific ECP (Docker Swarm and Kubernetes). We decided against TOSCA because of its tool-chain complexity and its tendency to cover all layers of abstraction (especially the infrastructure layer). Accordingly, we identified the need for creating a new DSL (at least for our research activities). Furthermore, to define a new DSL provides the maximum flexibility in covering all of the mentioned and derived requirements.

\begin{figure}[b]
\begin{center}
\includegraphics[width=.5 \textwidth]{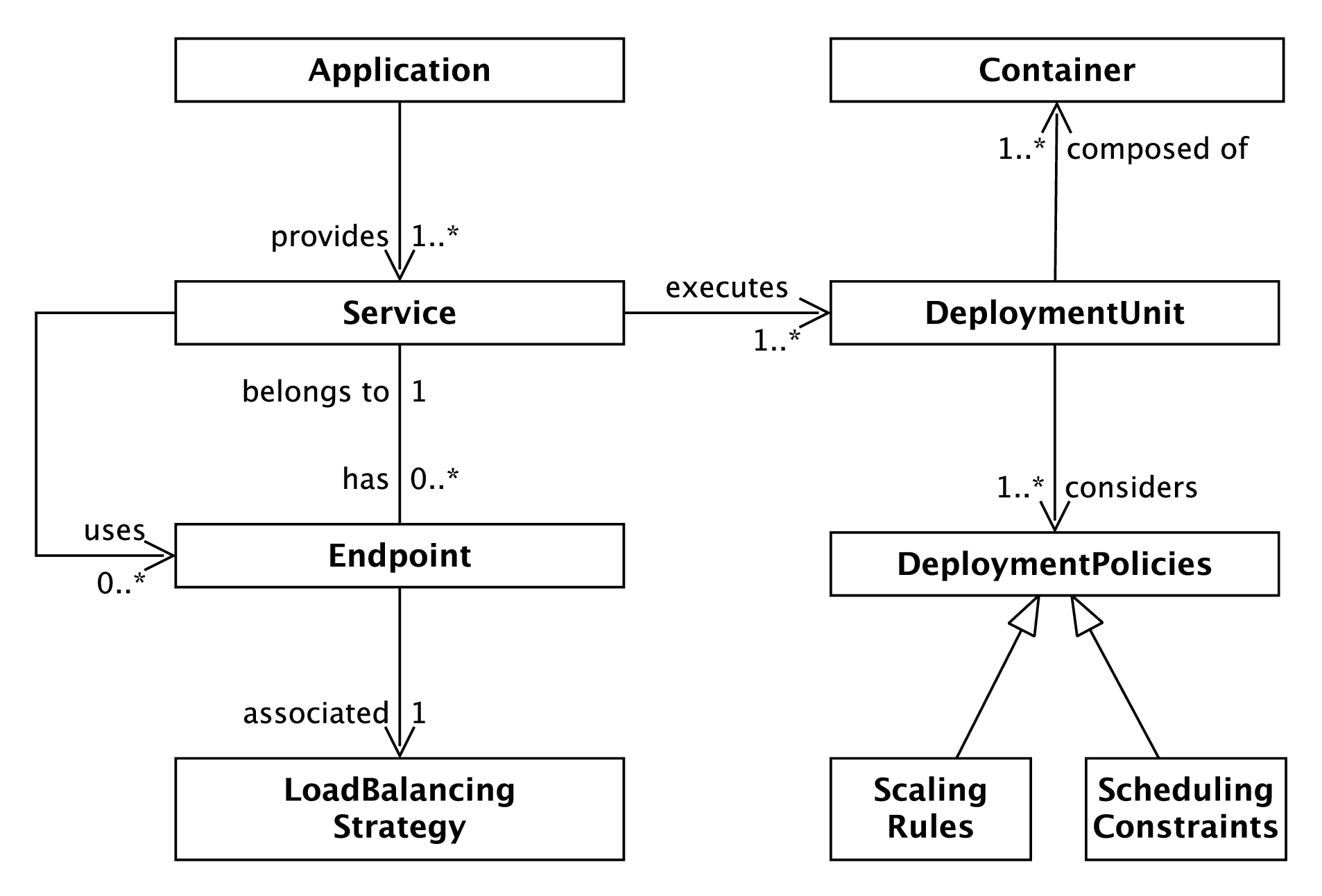}
\caption{DSL Core Language Model} 
\label{fig:DSL_corelangeuagemodel}
\end{center}
\end{figure}

Figure \ref{fig:DSL_corelangeuagemodel} summarizes the core language model for the resulting DSL. An \textbf{Application} \textit{provides} a set of \textbf{Service}s. A \textbf{Service} can have an \textbf{Endpoint} on which its features are exposed. One \textbf{Endpoint}  \textit{belongs} exactly to one \textbf{Service} and is \textit{associated} with a \textbf{Load Balancing Strategy}. A \textbf{Service} can \textit{use} other \textbf{Endpoint}s of other \textbf{Service}s as well. These \textbf{Service}s can be external Services that are not part of the application deployment itself. However, each internal \textbf{Service} \textit{executes} at least one \textbf{DeploymentUnit} which is \textit{composed of} one or more \textbf{Container}s. Furthermore, schedulers of ECPs should \textit{consider} \textbf{DeploymentPolicies} for  \textbf{DeploymentUnit}s. Such \textbf{DeploymentPolicies} can be workload considering \textbf{Scaling Rules} but also general \textbf{Scheduling Constraints}. 

Table \ref{tab:dsl-concept-mapping} relates these DSL concepts to identified requirements of Section \ref{sec:dsl} and initially identified trends in containerization of Section \ref{sec:ecp_based_cna}. Multi-Cloud support (requirement R4) is not directly mapped to a specific DSL concept. It is basically supported by separating the description of the ECP \cite{Kratzke2017} and the description of the CNA. Therefore, multi-cloud support must not be a part of the CNA DSL itself, what makes the CNA description less complex. Multi-cloud support is simply delegated to the lower level ECP. This kind of multi-cloud handling for ECPs is explained in more details by  \cite{Kratzke2017}.
\begin{table*}[t]
	\caption{Mapping DSL concepts to derived requirements (R1-R6) and containerization trends (AD, SD, DU, ..., CL).}
	\label{tab:dsl-concept-mapping}
	\footnotesize
	\centering
	\begin{tabular}{lcccccc|ccccccc}
		Concept                               & R1 & R2 & R3 & R4 & R5 & R6 & AD & SD & DU & SCHED & LB & AS & CL \\
		\hline
		Application                           &      &       &      &      &  x   &       & x  &       &        &              &      &       &      \\
		Service                                 &  x   &       &      &      &       &       &    &   x   &        &              &      &       &  x   \\
		Endpoint                               &      &       &  x   &       &       &       &    &   x   &        &              & x   &       &  x   \\
		DeploymentUnit                   &  x   &       &      &       &       &       &    &       & x       &   x        &      &       &  x   \\
		Container                             &  x   &       &      &       &       &       &    &       &  x      &   x        &      &       &    \\
		DeploymentPolicies             &       &       &      &  x     &       &  x   &    & x     &        &   x        &        &   x    &  x  \\
		LoadBalancingStrategy       &       &       &      &       &       &  x   &    & x     &        &           &   x   &   x    &    \\
		Scaling Rules                       &       &   x    &      &  x     &       &  x   &    & x     &        & x        &        &   x    &  x  \\
		Scheduling Constraints       &       &   x     &      &  x     &       &  x   &    & x     &        & x        &        &   x    & x   \\
	\end{tabular}
\end{table*}

According to \cite{van2000domain} we implemented this core language model as a declarative, internal DSL in Java. Although Java is a quite uncommon language to build a DSL, as a full purpose programming language it provides maximum flexibility to find DSL internal solutions. On the other hand, making use of proven software patterns makes it even possible to  provide human readable forms of application definitions (see Listing \ref{lst:demo}). To keep the description of a CNA simple to use and also short, we used the \textit{Builder Pattern} \cite{gamma2000design}. The usage of this pattern allows a flexible definition of a CNA without having to pay attention to the order of the description. The concrete syntax is shown exemplary using an example service as part of a \textit{SockShop} reference application that we used for our evaluation (Listing \ref{lst:demo}).

\section{\uppercase{Evaluation}}
\label{sec:evaluation}

 \begin{figure}[t]
	\begin{center}
		\includegraphics[width=.9 \columnwidth]{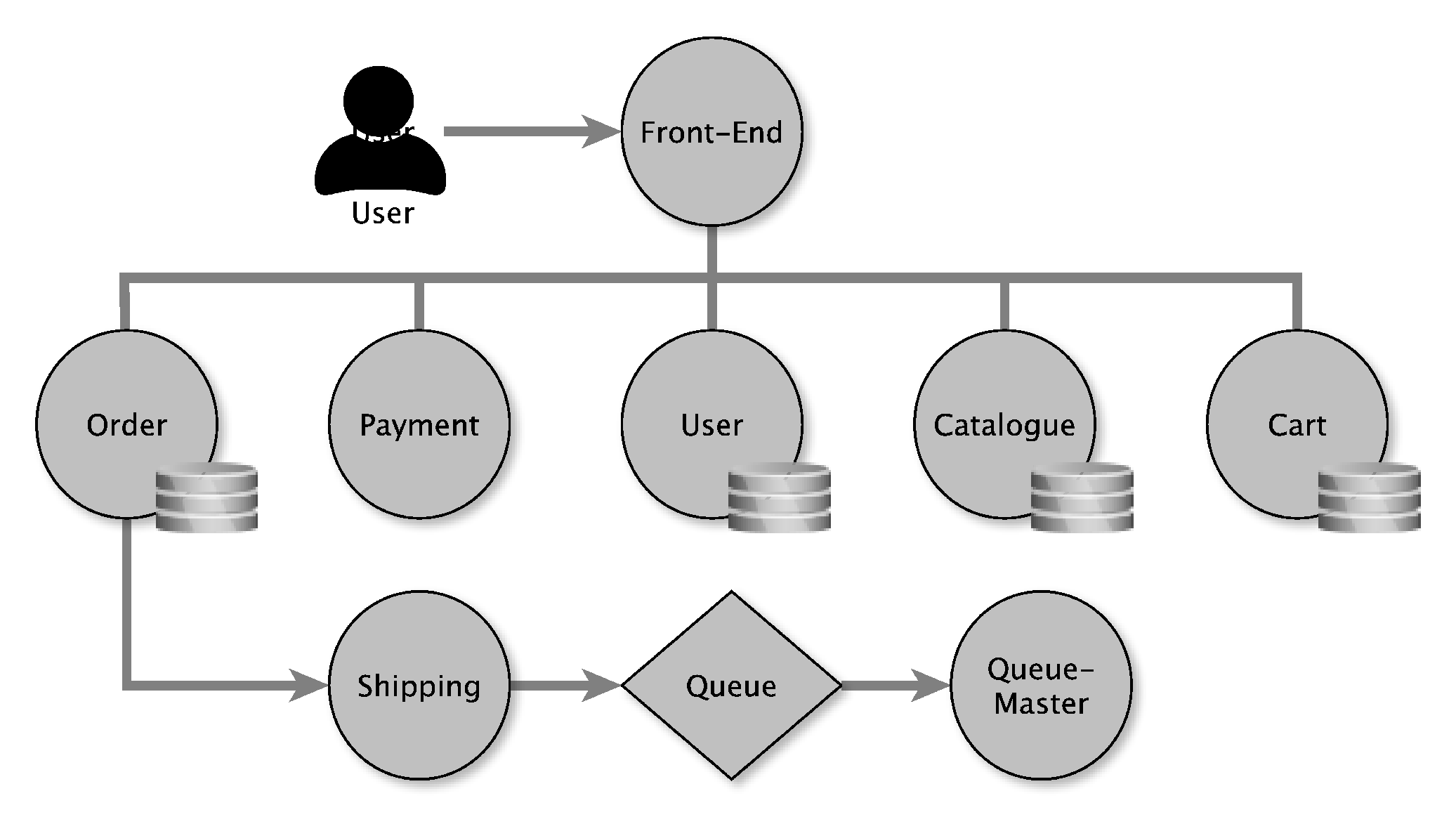}
		\caption{Architecture of the reference application \textit{Sock Shop}, according to \cite{sockshop}}
		\label{fig:sockshop}
	\end{center}
\end{figure}

\begin{figure*}[t]
	\centering
	\includegraphics[width=0.9\textwidth]{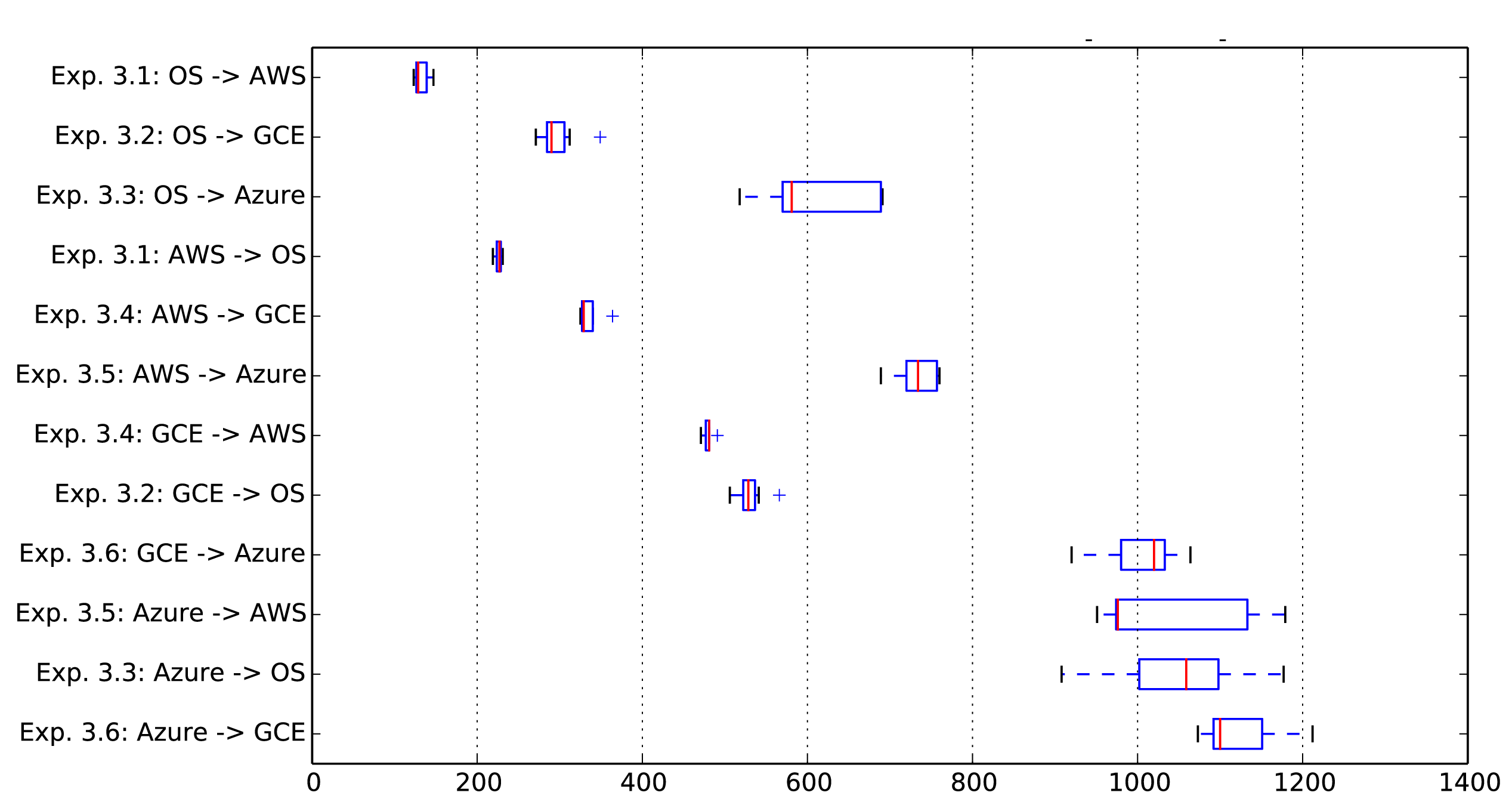}
	\caption{Measured durations of application migrations [seconds]}
	\label{fig:clustertransfer}
\end{figure*}

\begin{lstlisting}[
language=Java, %Damit das "new" nicht fett ist, so schlimm finde ich das nicht
float=*,
floatplacement=t,
xleftmargin=0.05\textwidth,
numbers=left,
caption=The payment service of the Sock Shop reference application expressed in the proposed DSL,
label=lst:demo,
basicstyle=\footnotesize\ttfamily]
DeploymentPolicy dPolicy = new DeploymentPolicy.Builder()
.rule(DeploymentPolicy.Type.NUMBER, 3)
.rule(DeploymentPolicy.Type.SELECTOR, "openStack.dc1")
.build();                
Container paymentContainer = new Container.Builder("payment")
.image("weaveworksdemos/payment:0.4.3")
.port(new Endpoint.Builder().containerPort(80).build())
.build();        
DeploymentUnit deploymentUnit = new DeploymentUnit.Builder("payment")
.container(paymentContainer)
.tag("app", "nginx")
.deploymentPolicy(dPolicy)
.build();   
Service service = new Service.Builder("payment")
.deploymentUnit(deploymentUnit)
.port(new Port.Builder("http")
.protocol(Port.Protocol.TCP).containerPort(80).targetPort(80).build())
.build();

new Generator.Builder().targetECP(Generator.ECP_TYPES.KUBERNETES)
.deyploment(service)
.build()
.write(new File("/path/to/folder"));
\end{lstlisting}

\noindent We validated that our DSL fulfills all requirements we defined in Section \ref{sec:dsl} by three evaluation steps:

\textbf{E1}. To evaluate the usability of the DSL for describing a containerized (\textbf{R1}) , auto-scalable (\textbf{R2}) deployment in a pragmatic way (\textbf{R3}), we described a microservice demonstration application. Therefore we selected \textit{Sock Shop}, a reference microservice e-commerce application for demonstrating and testing of microservice and cloud-native technologies \cite{sockshop}. Sock Shop is developed using technologies like Node.js, Go, Spring Boot and MongoDB and is one of the most complete reference applications for cloud-native application research according to \cite{aderaldo2017benchmark}. As shown in Figure \ref{fig:sockshop}, the application consists of nine services. Due to page limitations, we only provide one description of the payment-service as example in Listing \ref{lst:demo}.

\textbf{E2}. To evaluate multi-cloud-support (\textbf{R4}) and ECP independence (part of \textbf{R5}) we deployed and operated the \textit{Sock Shop} on two ECPs hosted on several IaaS infrastructures. As type representatives we selected Docker Swarm Mode (Version 17.06) and Kubernetes (Version 1.7). The ECPs consist of five working machines (and one master) hosted on the IaaS infrastructures OpenStack, Amazon AWS, Google GCE and Microsoft Azure. 

\textbf{E3}. For demonstrating IaaS independence (\textbf{R5}) we migrated the deployment between various IaaS infrastructures of Amazon Web Services, Microsoft Azure, Google Compute Engine and a research institution specific OpenStack installation. To validate all migration possibilities we have done the following experiments:

\begin{footnotesize}
\begin{itemize}
	\item \textbf{E3.1}: Migration OpenStack\footnote{\textit{Own Plattform}, machines with 2vCPUs} $\Leftrightarrow$ AWS \footnote{Region \textit{eu-west-1}, Worker node type \textit{m4.xlarge}} 
	\item \textbf{E3.2:} Migration OpenStack $\Leftrightarrow$ GCE \footnote{Region \textit{europe-west1}, Worker node type \textit{n1-standard-2}} 
	\item \textbf{E3.3}: Migration OpenStack $\Leftrightarrow$ Azure \footnote{Region \textit{europewest}, Worker node type \textit{Standard\_A2}} 
	\item \textbf{E3.4}: Migration $\Leftrightarrow$ and GCE
	\item \textbf{E3.5}: Migration AWS $\Leftrightarrow$ Azure
	\item \textbf{E3.6}: Migration GCE $\Leftrightarrow$ Azure
\end{itemize}
\end{footnotesize}

We have used Kubernetes and Docker Swarm\footnote{Due to page limitations we only present Kubernetes data. However, our experiments revealed that most runtime is spent in infrastructure specific handling and not due to the choice of the elastic container platform.} as ECP for the Sock Shop deployment. Every experiment is a set of migrations in both directions. E.g., evaluation experiment E3.1 includes migrations from OpenStack to AWS and from AWS to OpenStack. All migrations with OpenStack as source or target infrastructure (E3.1-E3.3) have been carried out ten times, all other (E3.4-E3.6) five times. 
The transfer times of the infrastructure migrations are shown in Figure \ref{fig:clustertransfer}. As the reader can see, the needed time for a infrastructure migration stretches from 3 minutes  (E3.1 OpenStack $\Rightarrow$ AWS) to more than 18min (E3.6 Azure $\Rightarrow$ AWS). Moreover, the transfer time for migrating also depends on the transfer direction between the \textit{source} and the \textit{target} infrastructure. E.g., as seen in E3.3, the migration Azure $\Rightarrow$ AWS takes four times longer than the  
reversed migration AWS $\Rightarrow$ Azure. Our analysis turned out, that the differences in the transfer times are mainly due to different blocking behavior of the IaaS API operations of different providers. Especially providers whose terminating operation of virtual machines or security groups are blocking operations show significantly longer reaction times. E.g., IaaS terminating operations of GCE and Azure wait until an operation is finished completely before starting the next one. This takes obviously longer than just waiting for the confirmation that an infrastructure operation has started (IaaS API behavior of OpenStack and AWS). However, and in all cases the reference application could be transferred completely and without downtime between all mentioned providers. The differences in transfer times are due to different involved IaaS cloud service providers and not due to the presented DSL.

\textbf{Limitations and Critical Discussion.}
In our current work we have not evaluated the migration of a stateful applications deployment with a mass of data. This would involve the usage of a storage cluster like Ceph or GlusterFS. The transfer of such kind of storage clusters will be investigated separately. We also rated the DSL pragmatism and practitioner acceptance higher than the richness of possible DSL expressions. This was a result according to discussions with practitioners \cite{KP2016}. This results in some limitations. For instance, our DSL is intentionally designed for container and microservice architectures, but has limitations to express applications out of this scope. This limits language complexity but reduces possible use cases. For applications outside the scope of microservice architectures, we recommend to follow more general TOSCA or CAMEL based approaches.

\section{\uppercase{Conclusion}}
\label{sec:conclusion}

\noindent Open issues in deploying cloud-native applications to  cloud infrastructures come along with the combination of multi-cloud interoperability, application topology definition/composition and elastic runtime adaption. This combination is -- to the best of the authors' knowledge -- not solved satisfactorily so far, because these three problems are often seen in isolation. It seems that cloud engineers (and researchers as well) just trust in picking only two out of these three options. Therefore, this paper strived for a more integrated point of view to overcome the observable isolation of these mentioned engineering and research trends \cite{KQ2017}. The key idea is to describe the platform independently from the application.  According to our lessons learned, the \textbf{infrastructure aware} deployment and operation of ECPs should be separated from \textbf{infrastructure and platform agnostic} deployment of applications. 

This paper focused on DSL design for the application level. However, if we take further research for the ECP and infrastructure level into consideration \cite{Kratzke2017}, we are able to demonstrate that a cloud-native application can be defined in a descriptive and infrastructure and platform-agnostic way simply using a specialized DSL. Our reference application composed of nine services could be expressed using the proposed prototypic version of such kind of a DSL. Furthermore, the application could be transferred between different cloud infrastructures within minutes and without downtimes.

Our DSL core language is implemented as internal DSL in Java to fulfill our own special demands in a fast and pragmatic way. But we see the need for a representation of our core language model without the overhead of a full purpose language like Java.  Further research will investigate whether it is useful to make use of more established topology DSLs like TOSCA and how to realize a comparable expressiveness like CAMEL. However, we do not strive for the technological possible, but also considere the balance between language expressiveness, pragmatism, complexity and practitioner acceptance.
\section*{\uppercase{Acknowledgements}}
 \noindent This research is funded by German Federal Ministry of Education and Research  (13FH021PX4). 
 Let us thank all the anonymous reviewers and their comments that improved this paper.

\bibliographystyle{apalike}
{\small
\bibliography{references}}

\end{document}